\documentclass[12pt]{article}
\usepackage[tbtags]{amsmath}
\usepackage{amsmath}
\usepackage{amssymb}
\usepackage[latin5]{inputenc}
\usepackage{amsthm}
\usepackage{graphicx}
\numberwithin{equation}{section}

\newcommand{\be}{\begin{equation}}
\newcommand{\ee}{\end{equation}}
\begin{document}
\vskip 2.0cm

 \title{\bf On the Classification of Fifth Order Quasi-linear Non-constant
Separant Scalar Evolution Equations of the KdV-type}
\author {
              {\bf  G\"{u}lcan \"{O}ZKUM$^1$ and Ay\c se H\"{u}meyra B\.ILGE$^2$ }\\
        {\it  $^1$Science and Letter Faculty, Kocaeli University,}\\
{\it Umuttepe Campus, Kocaeli, Turkey}\\
 {\it  $^2$Faculty of Sciences and Letters, Kadir Has University}\\
 {\it  Istanbul, Turkey}\\
        {\it  e-mail: ozkumg@itu.edu.tr and ayse.bilge@khas.edu.tr}
          }
\maketitle
\begin{abstract}

Fifth order, quasi-linear, non-constant separant evolution
equations are of the form $u_t=A\frac{\partial^5 u}{\partial
x^5}+\tilde{B}$, where $A$ and $\tilde{B}$ are functions of $x$,
$t$, $u$ and of the derivatives of $u$ with respect to $x$ up to
order $4$.
 We use the existence of a ``formal symmetry", hence the existence
of ``canonical conservation laws" $\rho_{(i)}$, $i=-1,\dots,5$ as
an integrability test. We define an evolution equation to be of
the KdV-Type, if all odd numbered canonical conserved densities
are nontrivial. We prove that fifth order, quasi-linear,
non-constant separant evolution equations of KdV type are
polynomial in the function $a=A^{1/5}$; $a=(\alpha u_3^2 +\beta
u_3+\gamma)^{-1/2}$, where $\alpha$, $\beta$ and $\gamma$ are
functions of $x$, $t$, $u$ and of the derivatives of $u$ with
respect to $x$ up to order $2$. We determine the $u_2$ dependency
of $a$ in terms of $P=4\alpha\gamma-\beta^2>0$ and we give an
explicit solution, showing that there are integrable fifth order
non-polynomial evolution equations.

\end{abstract}

\baselineskip=16pt

\section{Introduction}

In the literature, the term ``integrable equations" refers to
nonlinear equations that can either be transformed to a linear
equation or solved by an inverse spectral transformation. The
methods to determine whether a given equation is likely to be
integrable or not are called ``integrability tests". An extensive
review of the integrability tests and a comprehensive list of
integrable equation have been given in \cite{Mikhailov1991}.

The Korteweg-deVries
 (KdV), Sawada-Kotera and Kaup
hierarchies are well known hierarchies of integrable equations;
the KdV hierarchy appears at all odd orders, while the
Sawada-Kotera and Kaup hierarchies exist at odd orders that are
not multiples of $3$. Most of the constant separant third order
integrable equations listed in \cite{Mikhailov1991} are
transformable to an equation belonging to one of these
hierarchies; it is in fact conjectured that the only exception is
the Krichever-Novikov equation \cite{Heredero1995}. The search for
new hierarchies of integrable equations starting at order seven
didn't also give any positive result \cite{Bilge1992}; this
situation was clarified for the polynomial and scale invariant
evolution equations by the remarkable result of Wang and Sanders
\cite{Sanders1998}, stating that in the class of polynomial scale
invariant equations, any integrable equation of order $7$ or
higher is a symmetry of a lower order equation. Thus, the open
problems were reduced to the classifications at the third and
fifth orders and to the classification of higher order
non-polynomial equations.

In a series of papers we have applied the ``formal symmetry"
method of \cite{Mikhailov1991} to generic, non-polynomial
equations; we have first proved that integrable evolution
equations of order $7$ and greater are quasi-linear
\cite{Bilge2005}; then we have shown that they are polynomial in
top three derivatives \cite{Mizrahi2009}. Motivated by these
polynomiality results, we decided to investigate the structure of
fifth order quasi-linear integrable equations with non-constant
separant to see whether they would be polynomial and transformable
to the constant separant case classified in \cite{Mikhailov1991}.

In the present work we consider quasi-linear fifth order evolution
equations $u_t=A\frac{\partial^5 u}{\partial x^5}+\tilde{B}$ and
we focus on the case where $\rho_{(3)}$ is nontrivial, a case
naturally distinguished by the classification algorithm using
canonical conserved densities. Evolution equations with an
unbroken sequence of canonical densities will be called of
``Korteweg-deVries (KdV)-Type" as opposed to the Sawada-Kotera and
Kaup equations \cite{Kaup1980} for which $\rho_{(3)}$ is trivial.

We introduce our notation and give basic definitions in Section 2.
The dependency on $u_4$ is easily determined as presented in
Section 3.1 and it is shown that $u_t$ is quadratic in $u_4$.  The
form of the dependency on $u_3$ depends on whether the canonical
conserved density $\rho_{(3)}$ is trivial or not. In Section 3.2.1
we show that for the KdV-Type equations, the non-triviality of
$\rho_{(3)}$ implies that $A^{1/5}=a=(\alpha u_3^2+\beta
u_3+\gamma)^{-1/2}$, where $\alpha$, $\beta$ and $\gamma$ are
certain functions of $x$, $t$, $u$, $u_1$ and $u_2$. We use the
conserved density conditions to determine the form of the
evolution equation; in particular, in Section 3.2.3, we show that
they are polynomial in $a$ and in $u_3$, by proving that the
coefficients of logarithmic and other transcendental functions of
$a$ vanish by virtue of the partial differential equations
satisfied by $\alpha$, $\beta$ and $\gamma$. The discriminant of
the quadratic expression $\alpha u_3^2+\beta u_3+\gamma$ is
denoted by $-P$ and assumed to be negative, to avoid real roots.
The $u_2$ dependency of the functions  $\alpha$, $\beta$ and
$\gamma$ are determined in terms of $P$ but the $u_1$, $u$, $x$
and $t$ dependencies could'nt be obtained. In Section 3.2.4, we
present a fifth order integrable equation, assuming that $u_t$ has
no $x$, $t$, $u$, $u_1$ dependency, hence we present a new
integrable equation in this class.

As discussed in Section 4, these results show that at the fifth
order there are non-constant separant integrable equations which
are possibly higher order analogues of the essentially nonlinear
class of equations at the third order \cite{Heredero1994}.
Furthermore, as part of ongoing work, we have shown that there are
candidates of integrable equations of the form
$u_t=a^{2k+1}u_{2k+1}+\dots$ at orders $k=3,4,5,6$, which supports
the existence of a new hierarchy, which nevertheless is expected
to be transformable to the KdV equation by a Miura type
transformation involving third order derivatives.

\section{Preliminaries}
\subsection{Notation}

We shall work with evolution equations in one space dimension $x$
and use the following notation for the partial derivatives of the
dependent variable $u$.
$$u_t=\frac{\partial u}{\partial t},\quad
u_{1}=\frac{\partial u}{\partial x}=u_x,\quad
u_k=u_{\underbrace{x...x}_{k-times}}=\frac{\partial^{k}
u}{\partial x^{k}}.$$ In the following we shall assume that all
functions are sufficiently differentiable. Infinitely
differentiable functions of $x$, $t$, $u$ and of the derivatives
of $u$ up to an arbitrary but finite order are called
``differential functions" \cite{Olver1986}. We shall use
subscripts for partial derivatives with respect to the derivatives
of $u$, as below
$$\varphi_k=\frac{\partial\varphi}{\partial u_{k}},\quad \varphi_{kj}=\frac{\partial^2\varphi}{\partial u_{k}\partial u_j}.$$
The labels of functions will be denoted by indices in parenthesis,
such as $\sigma_{(i)}$ or $\rho_{(i)}$.

A scalar evolution equation of order $m$ is of the form
$$u_t=F(x,t,u,u_1,\dots,u_m). \eqno(1)$$
The total derivative operators with respect to $x$ and $t$ are
respectively denoted by $D$ and $D_t$. That is, if $\varphi$ is a
differentiable function of $x$, $t$, $u$ and the partial
derivatives of $u$ with respect to $x$ up to order $k$, then
$$D\varphi=\frac{\partial \varphi}{\partial x} +\frac{\partial
\varphi}{\partial u}\,u_{1}+\frac{\partial \varphi}{\partial
u_{1}}\,u_{2}+...+\frac{\partial \varphi}{\partial u_{k}}\,u_{k+1}
=\varphi_x+\varphi_0u_1+\dots+\varphi_k u_{k+1}.\eqno(2)$$ If
$u(x,t)$ satisfies (1) and $\varphi$ is as above, then
$$ D_t\varphi=\frac{\partial \varphi}{\partial t}
+\frac{\partial \varphi}{\partial u}\,F+\frac{\partial
\varphi}{\partial u_{1}}\,DF+...+\frac{\partial \varphi}{\partial
u_{k}}\,D^kF =\varphi_t+\varphi_0F+\dots+\varphi_k D^k
F.\eqno(3)$$

\subsection{The ``Level Grading"}

The notions of ``level grading" and level homogeneity introduced
in \cite{Mizrahi2009} are analogues of scaling and scale
homogeneity for polynomials; in the level grading context, the
``level above a base level $k$" reflects the number of
differentiations applied to a function that depends on the
derivatives of order at most $k$. The crucial property of the
level grading is its invariance under integrations by parts. This
property implies that ``top level" terms of a conserved density
$\rho$ will give top level terms in its time derivative $D_t\rho$,
hence we may omit lower level terms and still track correctly
higher level integrability conditions. The ``level of $u_n$ above
the base level $k$" is defined as $n-k$ and clearly it is equal to
the number of differentiations above $k$. The level of a monomial
$u_{n_1}^{b_1}\dots u_{n_m}^{b_m}$ is then defined as $b_1
(n_1-k)+\dots +b_m(n_m-k)$. For example, if the base level is
$k=4$, $u_5$ has level $1$, while if the base level is $k=3$,
$u_5$ and $u_4^2$ are both monomials of level $2$, hence, for
example, the right hand side of (14) is a sum of level homogeneous
expressions of levels $2$, $1$ and $0$ above the base level $k=3$.
In the present work, our results are based on explicit
computations including derivatives of all orders, however we still
indicate level related results whenever necessary.

\subsection{Formal Symmetries}

The Frechet derivative of a differential function $F$ is defined
by
$$F_{*}= \sum_{i=0}^{m}\;{\frac{\partial F}{\partial u_{i}}\;D^{i}}=\sum_{i=0}^{m}\;F_i D^{i}.\eqno(4)$$
Let  $\sigma=\sigma(x,t,u,...,u_{n})$ and
$\rho=\rho(x,t,u,...,u_{n})$ be differential functions. If
$$\sigma_{t}=F_{*}\sigma, \eqno(5)$$
then $\sigma$ is called a ``symmetry". If one can find a
differential function $\eta$ such that
$$\rho_{t}=D\eta\eqno(6)$$
then $\rho$ is called a ``conserved density". Let $R$ be a linear
operator. If for any symmetry $\sigma$, $R\sigma$ is also a
symmetry then $R$ is called a ``recursion operator". If $R\sigma$
is a symmetry, then the equation
$$(R\sigma)_t=F_*(R\sigma)\eqno(7)$$ gives
$(R\sigma)_{t}=R_{t}\sigma + RF_{*}\sigma= F_{*}R\sigma,$ hence
$$(R_t+[R,F_*])\sigma=0.\eqno(8)$$
Therefore, if $R$ satisfies the operator equation (8), it
certainly satisfies (7). Note however that in this case there is
no guarantee that the quantity $R\sigma$ which satisfies the
symmetry equation (7) is a local function. This suggests that if
we would solve the operator equation (8), we could as well work
with a pseudo-differential operator $R$. In addition, we may only
be interested in solving the operator equation (8) up to a finite
order. A pseudo-differential operator that solves (8) up to a
finite order is called a ``formal symmetry". The existence of a
formal symmetry has been proposed as an integrability test in
\cite{Mikhailov1991}. This approach has the advantage that the
existence of the recursion operator, hence the classification of
integrable equations is insensitive to the order and the
functional form of the operator $R$.  We have used the symbolic
programming language REDUCE in order to compute the partial
differential equations implied by these conserved density
conditions.

\section {The form of integrable equations}

The criterion of integrability  by the existence of a formal
symmetry \cite{Mikhailov1991} implies the existence of certain
conserved densities $\rho_{(i)}$, $i=-1,0,1,...$. It is well known
that for any order $m$, if $F_{m} =\frac{\partial F}{\partial
u_{m}},\quad  F_{m-1} =\frac{\partial F}{\partial u_{m-1}}$ then
$\rho_{(-1)}=F_{m}^{-1/m}\quad {\rm and}  \quad
\rho_{(0)}=F_{m-1}/F_{m}$ are always conserved densities. The
expression of the conserved densities $\rho_{(i)}$, $i=1,\dots, 5$
are given in Appendix A\cite{Ozkum2010}.

\subsection{Dependency on $u_4$}

We assume that the evolution equation is of the form
$$u_{t}=a^5u_{5}+\tilde{B},\eqno(9)$$
where $a$ and $\tilde{B}$ are functions of $x$, $t$, $u$, $u_i$,
$i=1,\dots,4$. Substituting (9) in the conserved densities given
in Appendix A, we can obtain explicitly the form of the canonical
conserved densities $\rho_{(1)}$ and $\rho_{(2)}$. However the
conserved densities $\rho_{(3)}$, $\rho_{(4)}$ and $\rho_{(5)}$
involve the expressions $\sigma_{(-1)}$, $\sigma_{(0)}$ and
$\sigma_{(1)}$ defined by $D_t\rho_{(i)}=D\sigma_{(i)}$, hence
they can be explicitly determined only after the lower order
conserved density conditions are solved. Note that, since
$\sigma_{(3)}$, $\sigma_{(4)}$ and $\sigma_{(5)}$ are not used
anywhere else, we may omit the total derivatives in $\rho_{(i)}$,
for $i=3,4,5$. In practice we solve the lower order conserved
density conditions up to a certain order, replace the results in
the other conserved densities and proceed iteratively. After
substitutions and integrations by parts, the conserved densities
are of the form given below.
\begin{eqnarray*}
\rho_{(-1)}&=&a^{-1},\\
\rho_{(0)}&=& 5a^4a_4u_5+\tilde{B}_4,\\
\rho_{(1)}&=&\tilde P_{(1)} u_5^2   +\tilde Q_{(1)} u_5   + \tilde R_{(1)},\\
\rho_{(2)}&=&\tilde M_{(2)} u_5^3   +\tilde P_{(2)} u_5^2 +\tilde Q_{(2)} u_5   + \tilde R_{(2)},\\
\rho_{(3)}&=&\tilde P_{(3)} u_6^2   +\tilde Q_{(3)} u_5^4 +\tilde
R_{(3)}u_5^3  +\tilde S_{(3)}u_5^2 +\tilde T_{(3)}u_5
+\tilde U_{(3)},\\
\rho_{(4)}&=&\tilde M_{(4)} u_6^2u_5+\tilde P_{(4)} u_6^2 +\tilde
Q_{(4)} u_5^5 +\tilde R_{(4)}u_5^4 +\tilde S_{(4)}u_5^3+\tilde T_{(4)}u_5^2 +\tilde U_{(4)}u_5\\
&&+\tilde V_{(4)},\\
\rho_{(5)}&=&\tilde P_{(5)} u_7^2   +\tilde Q_{(5)} u_6^3 +\tilde
R_{(5)} u_6^2u_5+ \tilde S_{(5)} u_6^2  +\tilde T_{(5)} u_5^5
+\tilde U_{(5)}u_5^4  +\tilde V_{(5)}u_5^3 \\
&&+\tilde W_{(5)}u_5^2+\tilde X_{(5)}u_5+\tilde
Y_{(5)}.\quad\qquad\qquad\qquad\qquad\qquad\qquad\qquad\qquad(10)
\end{eqnarray*}
In the equations above $\tilde{B}_4=\frac{\partial \tilde
B}{\partial u_4}$ and the coefficients $\tilde P_{(i)}$, $\tilde
Q_{(i)}$, $\tilde R_{(i)}$ ($i=1,\dots,5$),\; $\tilde S_{(i)}$,
$\tilde T_{(i)}$, $\tilde U_{(i)}$ ($i=3,4,5$), $\tilde M_{(2)}$,
$\tilde M_{(4)}$, $\tilde V_{(4)}$, $\tilde V_{(5)}$, $\tilde
W_{(5)}$, $\tilde X_{(5)}$ and $\tilde Y_{(5)}$ are functions of
$x$, $t$, $u$, $u_1$, $u_2$, $u_3$, $u_4$ that can be computed
from the expressions of the canonical densities given in the
Appendix A. We omit their explicit expressions except for
$$\tilde P_{(1)}=-\frac{a_4^2}{a},\quad \tilde P_{(3)}=\frac{3}{2}\,a[a_{44} a-4 a_4^2].\eqno(11)$$
In \cite{Bilge2005}, Corollary 4.3, it has been shown that
conserved densities of order $n$ larger than the order of the
evolution equation should be at most quadratic in their highest
derivatives; thus in particular, we should have $\tilde
M_{(2)}=0$.

The coefficient of $u_7^2 u_5$ in $D_t\rho_1$ gives
$$10 a_4 a^3 (a_{44} a-4 a_4^2)=0.\eqno(12)$$
Thus we should either have $a_4=0$, or if $a_4\ne 0$, then $a_{44}
a-4 a_4^2=0$. If $a_4=0$, then the coefficient of $u_5^2$ in
$D_t\rho_{(-1)}$ is an expression that involves $a$ and
$\tilde{B}_4$. Differentiating this expression with respect to
$u_4$ twice we obtain $\tilde B_{444}=0$.

On the other hand if $a_{44} a-4 a_4^2=0$, then, $\tilde
P_{(5)}=-20 a_4^2 a^3$ and $\tilde Q_{(5)}=\frac{880}{3}a_4^3a^2$
and the coefficient of $u_8^2u_6u_5$ gives $a_4=0$. When $a_4=0$,
in $\rho_{(4)}$, $\tilde M_{(4)}=\tilde P_{(4)}=\tilde
Q_{(4)}=\tilde R_{(4)}=0$. By the Corollary 4.3 of
\cite{Bilge2005}, $\rho_{(4)}$ should be quadratic in $u_5$, hence
$\tilde S_{(4)}$ should be zero. This condition gives again
$\tilde B_{444}=0$. Thus in either case, the $u_4$ dependency is
determined as below.
$$a_4=0,\quad \tilde{B}_{444}=0.\eqno(13)$$

\subsection{Dependency on $u_3$}

As a result of (13) the evolution equation reduces to the form
$$u_{t}=a^5u_{5}+Bu_{4}^{2}+Cu_{4}+G\eqno(14)$$
where $a$, $B$, $C$ and $G$ are now functions of $x$, $t$, $u$,
$u_1$, $u_2$ and $u_3$.

With this form of $F$ the canonical conserved densities reduce to
the form below.  Although the explicit expressions of the
conserved densities have been obtained, we use the generic
expressions given below for computational efficiency.
\begin{eqnarray*}
\rho_{(-1)}&=&a^{-1},\\
\rho_{(1)}&=&P_{(1)}u_{4}^2+Q_{(1)}u_{4}+R_{(1)},\\
\rho_{(3)}&=&(P_{(2)}u_{5}^2+Q_{(2)}u_{4}^{4})+R_{(2)}u_{4}^{3}+S_{(2)}u_{4}^{2}+T_{(2)}u_{4}+U_{2},\\
\rho_{(5)}&=&(P_{(3)}u_{6}^2+Q_{(3)}u_{5}^{3}+R_{(3)}u_{5}^{2}u_{4}^{2}+S_{(3)}u_{4}^{6})+(T_{(3)}u_{5}^{2}u_{4}+U_{(3)}u_{4}^{5})\\
          &&+(V_{(3)}u_{4}^{4}+W_{(3)}u_{5}^{2})+X_{(3)}u_{4}^{3}+Y_{(3)}u_{4}^{2}+Z_{(3)}u_{4}+M_{(3)}\;\;\qquad\qquad(15)
\end{eqnarray*}
The coefficients $P_{(i)}$, $Q_{(i)}$, $R_{(i)}$ ($i=1,2,3$),\;
$S_{(i)}$, $T_{(i)}$, $U_{(i)}$ ($i=2,3$),\; $V_{(3)}$, $W_{(3)}$,
$X_{(3)}$, $Y_{(3)}$, $Z_{(3)}$ and $M_{(3)}$ are functions of
$x$, $t$, $u$, $u_1$, $u_2$, $u_3$.

\noindent {\bf Definition 3.1.} An evolution equation
$u_t=F(x,t,u,\dots,u_m)$ of odd order $m$ is called of
``KdV-Type", if it admits an infinite sequence of non-trivial
conserved densities at all orders.

In terms of canonical densities the condition above means that all
the odd numbered canonical densities $\rho_{(2k+1)}$, in
particular, the canonical conserved density $\rho_{(3)}$ is
non-trivial. Note that these conserved densities are respectively
of levels $0$, $2$, $4$ and $6$ above the base level $3$, hence
their total derivatives with respect to time will be respectively
of levels $5$, $7$, $9$ and $11$.

We shall use the conditions obtained by equating to zero the
coefficients  of top $3$ levels terms in the conserved density
conditions as listed in Table 1. The coefficients of these
monomials denoted as $K_{ijk}$, where the first index $i$ denotes
the label of the conserved density, the second index $j$ denotes
the level of that term (as a hexadecimal number) and $k$ counts
the number of such conditions.  For example, $K_{3b4}$ is the
$4$th condition obtained as a coefficient of a term of level $11$
($b$ as hexadecimal) in $D_t\rho_{(3)}$.

\begin{table}
\caption{ The conserved density conditions for $\rho_{(i)}$
getting from the coefficients of terms of top $3$ levels.}
\qquad\qquad\qquad
\begin{tabular}{r l l l} \hline
$\rho_{(i)}$ & Top level & Top-1 level & Top-2 level \\
\hline  $\rho_{(-1)}$ & $u_{5}^2u_{4}$ - $K_{051}$, & $ u_{5}^2$
-$K_{041}$,& $u_{4}^{3}$ - $K_{031}$\\&
$u_{4}^{5}$ - $K_{052}$ & $u_{4}^{4}$ - $K_{042}$ & - \\

\hline $\rho_{(1)}$ & $u_{6}^2u_{4}$ - $K_{171}$, & $u_{6}^2$ -
$K_{161}$, & -\\& $u_{5}^3u_{4}$ - $K_{172}$, &$u_{5}^3$ - $K_{162}$, & -\\
& $u_{5}^2u_{4}^{3}$ - $K_{173}$, & $u_{5}^2u_{4}^{2}$ -
$K_{163}$, & $u_{5}^2u_{4}$ - $K_{151}$,\\&
$u_{4}^7$ - $K_{174}$&   $u_{4}^6$ - $K_{164}$, & $u_{4}^5$ - $K_{152}$\\

\hline $\rho_{(2)}$ & $u_{7}^2u_{4}$ - $K_{291}$,& $u_{7}^2$ - $K_{281}$, & -\\
&$u_{6}^3$ - $K_{292}$,& - & - \\& $u_{6}^2u_{5}u_{4}$ -
$K_{293}$, &$u_{6}^2u_{5}$ - $K_{282}$, & $u_{6}^2u_{4}$ -
$K_{271}$,\\ & $u_{6}^2u_{4}^{3}$ - $K_{294}$, &
$u_{6}^2u_{4}^{2}$ - $K_{283}$, & -\\& $u_{5}^4u_{4}$ - $K_{295}$,
& $u_{5}^4$ - $K_{284}$, & -
\\& $u_{5}^3u_{4}^{3}$ - $K_{296}$,& $u_{5}^3u_{4}^{2}$ - $K_{285}$, & $u_{5}^3u_{4}$ - $K_{272}$,\\&
$u_{5}^2u_{4}^{5}$ - $K_{297}$, &  $u_{5}^2u_{4}^{4}$ - $K_{286}$,
& $u_{5}^2u_{4}^{3}$ - $K_{273}$,\\&
$u_{4}^9$ - $K_{298}$ & $u_{4}^{8}$ - $K_{287}$ & $u_{4}^{7}$ - $K_{274}$\\

\hline $\rho_{(3)}$ & $u_{8}^2u_{4}$ - $K_{3b1}$, & $u_{8}^2$ -
$K_{3a1}$, & -\\& $u_{7}^2u_{6}$ - $K_{3b2}$, & $u_{7}^2u_{5}$ -
$K_{3a2}$,&-\\& $u_{7}^2u_{5}u_{4}$ - $K_{3b3}$,& - & - \\&
$u_{7}^2u_{4}^{3}$ - $K_{3b4}$, &$u_{7}^2u_{4}^{2}$ - $K_{3a3}$, &
$u_{7}^2u_{4}$ - $K_{391}$,\\& $u_{6}^3u_{5}$ - $K_{3b5}$,& - & -
\\& $u_{6}^3u_{4}^{2}$ - $K_{3b6}$, & $u_{6}^3u_{4}$ - $K_{3a4}$, &
$u_{6}^3$ - $K_{392}$ \\ & $u_{6}^2u_{5}^{2}u_{4}$ - $K_{3b7}$, &
$u_{6}^2u_{5}^{2}$ - $K_{3a5}$, & -\\& $u_{6}^2u_{5}u_{4}^{3}$ -
$K_{3b8}$, & $u_{6}^2u_{5}u_{4}^{2}$ - $K_{3a6}$,&
$u_{6}^2u_{5}u_{4}$ - $K_{393}$,\\& $u_{6}^2u_{4}^{5}$ -
$K_{3b9}$,& $u_{6}^2u_{4}^{4}$ - $K_{3a7}$,& $u_{6}^2u_{4}^{3}$ -
$K_{394}$,\\& $u_{5}^5u_{4}$ - $K_{3ba}$, & $u_{5}^5$ - $K_{3a8}$,
& -\\& $u_{5}^4u_{4}^{3}$ - $K_{3bb}$, & $u_{5}^4u_{4}^{2}$ -
$K_{3a9}$,& $u_{5}^4u_{4}$ - $K_{395}$,
\\& $u_{5}^3u_{4}^{5}$ - $K_{3bc}$, & $u_{5}^3u_{4}^{4}$ - $K_{3aa}$, & $u_{5}^3u_{4}^{3}$ - $K_{396}$,
\\ & $u_{5}^2u_{4}^{7}$ - $K_{3bd}$,& $u_{5}^2u_{4}^{6}$ - $K_{3ab}$, & $u_{5}^2u_{4}^{5}$ - $K_{397}$,  \\&
$u_{4}^{11}$ - $K_{3be}$, & $u_{4}^{10}$ - $K_{3ac}$, &  $u_{4}^{9}$ - $K_{398}$\\
\hline
\end{tabular}
\end{table}

\subsubsection{Top level computations}

We start by collecting those top level equations that will
determine $B$ and give a third order ordinary differential
equation for the $u_3$ derivative of $a$. Although the explicit
forms of the $P_{(i)}$'s are known, it will be more convenient to
consider them as new unknown functions.

From the condition $K_{051}$ we get
$$K_{051}:a_{333}a^{6} - 7a_{33}a_{3}a^{5} - \frac{4}{5}\;a_{33}aB + 8a_{3}^{3}a^{4}-\frac{8}{5}\;a_{3}^{2}B=0.\eqno(16)$$
Recall that as we are studying KdV-Type equations we are assuming
that $\rho_{(3)}$ is nontrivial, hence $P_{(2)}\neq 0$. We then
use the equations
\begin{eqnarray*}
\qquad\qquad\qquad K_{3b1}&:& 35a_{3}a^{4}P_{(3)} - 5(P_{(3)})_{3}\;a^{5} + 4B P_{(3)}=0,\\
K_{171}&:& 15a_{3}a^{4}P_{(1)} - 5(P_{(1)})_{3}\;a^{5} + 4B P_{(1)}=0,\\
K_{291}&:& 25a_{3}a^{4}P_{(2)} - 5(P_{(2)})_{3}\;a^{5} + 4B P_{(2)}=0,\\
K_{292}&:& 20a_{3}a^{4}P_{(2)} - 5(P_{(2)})_{3}\;a^{5} + 6B
P_{(2)}=0.\qquad\qquad\qquad(17)
\end{eqnarray*}
Taking the difference of the last two  equations, we solve $B$ as
$$B =\frac{5}{2}\;a_{3} a^{4}.\eqno(18)r$$
From the remaining equations in (17) we find that the derivatives
of $P_{(i)}$ and $a$ with respect to $u_3$ are proportional, and
we solve them as below
$$P_{(1)}= P_{(1o)}a^{5},\quad P_{(2)}= P_{(2o)}a^{7},\quad P_{(3)}= P_{(3o)}a^{9}.  \eqno(19)$$
where $P_{(io)}$ are functions of $x$, $t$, $u$, $u_1$ and $u_2$.
Finally, $K_{051}$ gives
$$ a_{333}a^{2} - 9a_{33}a_{3}a  + 12a_{3}^{3}=0.\eqno(20)$$
Assuming $a=Z^{k}$ and substituting in (20), we see that for
$k=1/2$, $Z_{333}=0$, hence $a$ is of the form
$$a=(\alpha {u_{3}}^2+\beta u_{3}+\gamma)^{-\frac{1}{2}}\eqno(21)$$
where $\alpha$, $\beta$ and $\gamma$ are functions of $x$, $t$,
$u$, $u_1$ and $u_2$. From (21) we can see that
$$a_3=-\frac{1}{2}\,a^3 (2\alpha u_3 +\beta),\quad a_i=-\frac{1}{2}\,a^3 (\alpha_i u_3^2  +\beta_i u_3+\gamma_i),\quad i=x,0,1,2.\eqno(22)$$
If we require $a$ to be real for every $u_{3}$, we see that
$$\beta^{2}-4\alpha\gamma<0,\quad {\rm and}\quad \alpha>0$$
and we define
$$P=4\alpha\gamma-\beta^{2}.\eqno(23)$$

From the expressions, $K_{293}$, $K_{3b2}$, $K_{3b3}$ and
$K_{3b8}$, we solve respectively $Q_{(2)}$, $Q_{(3)}$, $R_{(3)}$
and $S_{(3)}$ but we omit the explicit expressions here. The rest
of the top level conditions are  automatically satisfied.

\subsubsection{Top-1 level computations}

By computing the coefficients of $u_4^2$, $u_5^2$ and $u_6^2$
respectively in $\rho_{(1)}$, $\rho_{(3)}$ and $\rho_{(5)}$, we
find that
$$ P_{(1o)}=\frac{1}{8}P,\quad P_{(2o)}=-\frac{3}{8}P,\quad P_{(3o)}=\frac{5}{2}P.\eqno(24)$$
From the (Top-1) level condition $K_{281}$, $C$ is obtained as,
$$C =\frac{5}{2}\frac{DP}{P}\;a^{5}+ 5\;(Da - a_{3}u_{4})\;a^{4}.\eqno(25)$$
We can again solve $R_{(2)}$, $T_{(3)}$ and  $V_{(3)}$
respectively from $K_{282}$, $K_{3a2}$ and  $K_{3a6}$ but we omit
these expressions.

Then from $K_{162}$, the coefficients of $u_3^2$  we obtain
$$\alpha=\alpha_{(0)}P^2,\eqno(26)$$
where $\alpha_{(0)}=\alpha_{(0)}(u,u_{1})$, and the coefficient of
$u_{3}$ gives
$$ \frac{\partial \beta}{\partial u_{2}}-\frac{1}{2}\;\beta
\frac{\partial P}{\partial u_{2}} \frac{1}{P}= 3\alpha_{(0)}P\,
\Big(\frac{\partial P}{\partial u_{1}}u_{2} +\frac{\partial P}
{\partial u}u_{1} +\frac{\partial P} {\partial x}
\Big).\eqno(27)$$ The remain conditions of the (Top-1) level
conditions are satisfied automatically.

\subsubsection{Top-2 level computations}

All of the (Top-2) level conditions except for $K_{292}$ are
identically satisfied. From the condition $K_{292}$ we get a
second order ordinary differential equation for the $u_3$
derivative of $G$. The form of this equation suggests that we have
to multiply it by $a^{-1}$ in order to integrate it. After
substituting
$$(2\alpha u_3+\beta)=-2a_3 a^{-3},\eqno(28)$$
the resulting expression is
$$a^{-3}G_{33}-3a^{-4}a_3G_{3}
+\sum_{i=1}^4\left(\lambda_{(i)}u_3+\mu_{(i)}\right)a^{2i}=0,\eqno(29)$$
where the $\lambda_{(i)}$'s and $\mu_{(i)}$'s depend on $x$, $t$,
$u$, $u_1$ and $u_2$. In order to integrate this equation, we need
to know the integrals of various powers of $a$. Note that $a^{-2}$
is an irreducible  quadratic polynomial in $u_3$, hence the
integrals of $a^2 u_3$ and $a^2$ will involve logarithms and
arctangent functions. However, the coefficients of these terms
will be zero, hence, instead of writing these integrals
explicitly, we just indicate these as new functions. These
integration formulas are given below.
\begin{eqnarray*}
\qquad\qquad\qquad\quad \quad\int a\
du_3&=&\alpha^{-1/2}\ln\left[P^{-1/2}\left(2\alpha^{1/2}
a^{-1}+2\alpha u_3+\beta\right)\right],\\
\int a u_3 \ du_3&=&\frac{1}{\alpha a}-\frac{\beta}{2\alpha}\int a\ du_3,\\
\int a^2 \ du_3&=&\psi=2P^{-1/2}\tan^{-1}\left(P^{-1/2}(2\alpha u_3+\beta)\right),\\
\int a^2 u_3 \ du_3&=&\varphi=-\frac{1}{\alpha}\ln (a)-\frac{\beta}{2\alpha}\int a^2\ du_3,\\
\int a^3 \ du_3&=&\frac{2}{P}\left(2\alpha
u_3+\beta\right)a.\qquad\quad\quad\quad\quad\quad
\quad\quad\quad\quad(30)
\end{eqnarray*}

The integrals of $a^k$ and $a^ku_3$ for $k\ge 3$ are evaluated
using the iterative formulas given below.
$$\int a^{k} \ du_3=\frac{2}{k-2}\frac{1}{P}\,\left(2\alpha
u_3+\beta\right)\,a^{k-2}
+\frac{4(k-3)}{k-2}\frac{1}{P}\,\alpha\int a^{k-2} \
du_3,\eqno(31)$$
$$\int a^{k}u_3 \ du_3=\frac{1}{2\alpha}\,\Big(-\frac{k}{2}+1\Big)^{-1} a^{k-2} -\frac{\beta}{2
\alpha}\int a^{k} \ du_3. \eqno(32)$$ The resulting integral is of
the form
$$G_{3}a^{-3}
+\kappa_{(1)} \psi+\kappa_{(2)} \varphi +\sum_{i=1}^3
\left(\nu_{(i)}u_3+\eta_{(i)}\right)a^{2i}=\chi,\eqno(33)$$ where
the $\kappa_{(i)}$'s, $\nu_{(i)}$'s, $\eta_{(i)}$'s and $\chi$
depend on $x$, $t$, $u$, $u_1$ and $u_2$. The coefficients of the
functions $\psi$ and $\varphi$ defined in Eqn.(33) are zero, after
we substitute the equations for $\alpha$ and $\beta$, hence
actually
$$\kappa_{(1)}=\kappa_{(2)}=0.$$
Multiplying (33) by  $a^3$, we have
$$G_{3}=\chi a^3 - \sum_{i=1}^3 \left(\nu_{(i)}u_3+\eta_{(i)}\right) a^{2i+3}.\eqno(34)$$
Integrating one more time, $G$ is obtained as
$$G = \left(G_{(1)}u_3+G_{(2)}\right) a^7 + \left(G_{(3)}u_3+G_{(4)}\right) a^5+ \left(G_{(5)}u_3+G_{(6)}\right)a^3
+(G_{(7)}u_3+G_{(8)}) a + G_{(9)}.\eqno(35)$$ The coefficients are
found explicitly from $K_{292}$, but we omit these expressions. It
follows that for $P\ne 0$ integrable evolution equations with
nontrivial $\rho_{(3)}$ are of the form
$$
u_{t}=a^{5}u_{5}+\frac{5}{2}\;a_{3}
a^{4}u_{4}^{2}+\Big(\frac{5}{2}\frac{DP}{P}a^{5}+ 5\left(Da -
a_{3}u_{4}\right)a^{4}\Big)u_{4}+\left(G_{(1)}u_3+G_{(2)}\right)a^7$$
$$\qquad\;+\left(G_{(3)}u_3+G_{(4)}\right)a^5 + \left(G_{(5)}u_3+G_{(6)}\right)a^3
+\left(G_{(7)}u_3+G_{(8)}\right)a + G_{(9)},\eqno(36)
$$
where $a$ and $P$ are given respectively by (21) and (23), the
$G_{(i)}$, $i=1,\dots, 6$ are certain functions of $\alpha$,
$\beta$ and $\gamma$, $G_{(7)}=G_{(7o)}P$, where $G_{(7o)}$ is a
function of $x$, $t$, $u$ and $u_1$ and $G_{(8)}$, $G_{(9)}$ are
functions of $x$, $t$, $u$, $u_1$ and $u_2$.

\vskip 0.2cm \noindent{\bf Proposition 3.1.} Let
$u_t=a^5u_5+\tilde{B}$ be an evolution equation where $a$ and
$\tilde{B}$ depend on $x$, $t$, $u$, $u_i$, $i=1,\dots,4$. If the
canonical densities $\rho_{(i)}$, $i=-1,\dots ,5$ are conserved,
and $\frac{\partial^2 \rho_{(-1)}}{\partial u_3^2}$ and
$\frac{\partial^2 \rho_{(3)}}{\partial u_3^2}$ are nonzero, then
integrable evolution equations are of the form in (36), where $a$
and $P$ are given respectively by (21) and (23), the $G_{(i)}$,
$i=1,\dots, 6$ are certain functions of $\alpha$, $\beta$ and
$\gamma$, $G_{(7)}=G_{(7o)}P$, where $G_{(7o)}$ is a function of
$x$, $t$, $u$ and $u_1$ and $G_{(8)}$, $G_{(9)}$ are functions of
$x$, $t$, $u$, $u_1$ and $u_2$.

\subsubsection{A new solution}

The equation (36) obtained above gives the form of the candidates
for integrable fifth order quasi-linear equations. In this section
we assume that $a$ is independent of $x$, $u$ and $u_1$ and obtain
an equation for which all of the canonical densities $\rho_{(i)}$,
$i=-1,\dots,5$ are conserved.

\noindent{\bf Proposition 3.2.} Let $u_t=a^5u_5+\tilde{B}$ be an
evolution equation where $a$ and $\tilde{B}$ depend on $u_i$,
$i=2,\dots,4$. If the canonical densities $\rho_{(i)}$,
$i=-1,\dots ,5$ are conserved, and $\frac{\partial^2
\rho_{(-1)}}{\partial u_3^2}$ and $\frac{\partial^2
\rho_{(3)}}{\partial u_3^2}$ are nonzero, then
\begin{eqnarray*}
 u_{t}&=&a^{5}u_{5}+Bu_{4}^{2}+Cu_{4} +\left(G_{(1)}
a^7 +G_{(3)}a^5+ G_{(5)}a^3 + G_{(7)}a\right)u_3 +
G_{(9)},\;\,(37)
\end{eqnarray*}
where
\begin{eqnarray*}
&&a=(\alpha u_3^2 +\beta u_3 +\gamma)^{-1/2},\quad \alpha=\alpha_{(0)}P^2, \quad \beta=0,\quad\gamma=\frac{P+\beta^2}{4\alpha},\\
&&P=(\mu u_2^2+\nu u_2+\kappa)^{-1}, \quad q=4\mu\kappa-\nu^2,\\
&&B=-\frac{5}{2}\,\alpha_{(0)} P^2 a^7 u_3,\quad
C=\Big(-\frac{15}{8}\,\alpha_{(0)}^{-1}a^2+\frac{5}{2}P\Big)\left(2\mu u_2+\nu\right)a^5 u_3,\\
&&
G_{(1)}=\frac{45}{128}\,\alpha_{(0)}^{-3}\,P^{-3}\big(qP-4\mu\big),
\quad
G_{(3)}=\frac{27}{16}\,\alpha_{(0)}^{-2}\,P^{-2}\Big(-qP+\frac{40}{9}\mu\Big),\\
&&G_{(5)}=\alpha_{(0)}^{-1}\,P^{-1}\Big(qP-\frac{15}{2}\mu\Big),
\quad G_{(7)}=G_{(7o)}P, \quad G_{(9)}=G_{(9o)},
\end{eqnarray*}
 and  $\mu$, $\nu$, $\kappa$, $\alpha_{(0)}$,  $G_{(7o)}$ and $G_{(9o)}$ are constants.

It has been checked that all even canonical densities are trivial
and all odd canonical densities up to $\rho_{(5)}$ are conserved.
Thus the equation above is integrable in the sense of admitting a
formal symmetry \cite{Mikhailov1991}. We have observed that all
the even canonical densities up to $\rho_{(4)}$ are trivial while
the odd canonical densities are nontrivial.

\section{Discussion of the Results}

In the course of this study, the existence of non-polynomial
integrable equations came as surprise, because we were expecting
that the polynomiality in top three derivatives would be
generalized to lower orders. We then remarked that this class
seems to be related to the essentially nonlinear class of third
order equations \cite{Mikhailov1991},
$$u_t=H=\big(A_1u_3^2+A_2 u_3+A_3\big)^{-1/2} (2 A_1 u_3+A_2)+ A_4,$$
where $A_i=A_i(x,u,u_1,u_2)$. This similarity is suggested by the
fact that the conserved density $\rho_{(-1)}$ is the same for both
equations. Note that a scaling of the $A_i$'s result in a scaling
of $H$; replacing $A_i$ by $(2/P)^{1/2} A_i$, $i=1,2,3$ one can
obtain $$\frac{\partial H}{\partial u_3}=a^3.$$ Hence
$\rho_{(-1)}$ is the same for the essentially nonlinear third
order class and the fifth order quasi-linear non-constant separant
class.

Then, in order to see whether these equations were part of a
hierarchy, we studied the top level terms of level homogeneous odd
order equations up to order $13$; surprisingly we have seen that
at each order, these flows were polynomial in $u_k$, $k\ge 3$ of
the form $u_t=a^m u_m + \dots$, i.e, with separant $a^m$.

Finally we note that the classification of essentially nonlinear
class of equations is discussed in \cite{Heredero1994} where it is
suggested to use a formulation where the dependent variable is
replaced by the canonical density  $\rho_{(-1)}$. With this
method, the classification of the essentially nonlinear equations
is partly completed in \cite{Heredero2005}; in particular the
classification of the equations with $A_1=0$ corresponding to
$\alpha_{(0)}=0$ in our case is obtained, while the case $A_1\ne
0$ is still open.

\section*{Acknowledgment}
This work is based on the Ph.D. thesis of the first author and it
is partially supported by the Turkish Scientific and Technical
Research Council. We thank Professors Metin G\"urses and Jing Ping
Wang for valuable discussions; for their suggestions to
concentrate on specific solution (MG) and for pointing our recent
work by Heredero (JPW).

\section*{Appendix A. Canonical Densities}

\begin{eqnarray*}
\rho_{(-1)}&=&a^{-1},\qquad\qquad\qquad\qquad\qquad\qquad\qquad\qquad\qquad\qquad\qquad\qquad\quad (A.1)\\
\rho_{(0)}&=&F_4/F_5,\quad\qquad\qquad\qquad\qquad\qquad\qquad\qquad\qquad\qquad\qquad\qquad\;\;\;(A.2)\\
\rho_{(1)}&=&2\,(D^2a) - (Da)^2 a^{-1} + 2\,(Da)\,a^{-5}F_4 -
\frac{2}{5}\,(DF_4)\,a^{-4} + \frac{1}{5}\,a^{-4}F_3\\
&&-\frac{2}{25}\,a^{-9}{F_4}^2,\qquad\qquad\qquad\qquad\qquad\qquad\qquad\qquad\qquad\qquad\quad\;\,(A.3)
\end{eqnarray*}
\begin{eqnarray*}
\rho_{(2)}&=&-2\,(D^2a)\,a^{-4} F_4 + 12\,(Da)^2 a^{-5} F_4 -
4\,(Da)\,(DF_4)\,a^{-4} \\
&&+ 3\,(Da)\,a^{-4} F_3-\frac{12}{5}\,(Da)\,a^{-9} {F_4}^2 -
\frac{3}{5}\,(DF_3)\,a^{-3} + \frac{2}{5}\,(D^2F_4)\,a^{-3} \\
&&+ \frac{12}{25}\,(DF_4)\,a^{-8} F_4  + \frac{2}{5}\,a^{-3} F_2 -
\frac{6}{25}\,a^{-8} F_3 F_4 +
\frac{8}{125}\,a^{-13}{F_4}^3,\quad\quad\;(A.4)
\end{eqnarray*}
\begin{eqnarray*}
\rho_{(3)}&=&-\frac{8}{5}\,(D^4a)\,a^2 -
\frac{16}{5}\,(D^3a)\,(Da)\,a - (D^3a)\,a^{-3} F_4 \\
&&- \frac{12}{5}\,(D^2a)^2 a + \frac{12}{5}\,(D^2a)\,(Da)^2 +
2\,(D^2a)\,(Da)\,a^{-4}F_4 \\
&&- \frac{7}{5}\,(D^2a)\,(DF_4)\,a^{-3} - \frac{9}{5}\,(D^2a)\,
a^{-3}F_3 + \frac{28}{25}\,(D^2a)\, a^{-8}{F_4}^2
\\
&&-\frac{3}{5}\,(Da)^4 a^{-1} + 2\,(Da)^3 a^{-5}F_4 +
\frac{14}{5}\,(Da)^2\,(DF_4)\, a^{-4} \\
&&+ \frac{48}{5}\,(Da)^2\, a^{-4}F_3 -
\frac{371}{25}\,(Da)^2\,a^{-9}{F_4}^2 -
\frac{14}{5}\,(Da)\,(DF_3)\,a^{-3} \\
&&-\frac{7}{5}\,(Da)\,(D^2F_4)\,a^{-3} +
\frac{126}{25}\,(Da)\,(DF_4)\,a^{-8}F_4 + 3\,(Da)\,a^{-3}F_2 \\
&&- \frac{21}{5}\,(Da)\,a^{-8}F_3F_4 +
\frac{42}{25}\,(Da)\,a^{-13}{F_4}^3 - \frac{3}{5}\,(DF_2)\,a^{-2}
\\
&&+ \frac{1}{5}\,(D^2F_3)\,a^{-2} +
\frac{9}{25}\,(DF_3)\,a^{-7}F_4 + \frac{1}{5}\,(D^3F_4)\,a^{-2} \\
&&- \frac{4}{25}\,(D^2F_4)\,a^{-7}F_4 -
\frac{7}{25}\,(DF_4)^2\,a^{-7} + \frac{12}{25}\,(DF_4)\,a^{-7}F_3
\\
&&-\frac{42}{125}\,(DF_4)\,a^{-12}{F_4}^2-
\frac{3}{5}\,a^{-1}\sigma_{(-1)} + \frac{3}{5}\,a^{-2}F_1 -
\frac{6}{25}\,a^{-7}F_2F_4 \\
&&- \frac{3}{25}\,a^{-7}{F_3}^2 +
\frac{21}{125}\,a^{-12}F_3{F_4}^2 -
\frac{21}{625}a^{-17}{F_4}^4.\qquad\qquad\qquad\quad\;(A.5)
\end{eqnarray*}
\begin{eqnarray*}
\rho_{(4)}&=&\frac{4}{5}\,(D^4a)\,a^{-2}F_4 -
\frac{66}{5}\,(D^3a)\,(Da)\,a^{-3}F_4 +
\frac{16}{5}\,(D^3a)\,(DF_4)\,a^{-2} \\
&&- (D^3a)\,a^{-2}F_3 + \frac{4}{5}\,(D^3a)\,a^{-7}{F_4}^2 -
\frac{42}{5}\,
(D^2a)^2\,a^{-3}F_4 \\
&&+ \frac{318}{5}\,(D^2a)\,(Da)^2\,a^{-4}F_4 -\frac{138}{5}\,(D^2a)\,(Da)\,(DF_4)\,a^{-3} \\
&&- 8\,(D^2a)\,(Da)\,a^{-8}{F_4}^2
-\frac{6}{5}\,(D^2a)\,(DF_3)\,a^{-2} +
\frac{18}{5}\,(D^2a)\,(D^2F_4)\,a^{-2} \\
&&+ \frac{64}{25}\,(D^2a)\,(DF_4)\,a^{-7}F_4
 - \frac{6}{5}\,(D^2a)\,a^{-2}F_2 + \frac{28}{25}\,(D^2a)a^{-7}F_3F_4 \\
&&- \frac{44}{125}\,(D^2a)\,a^{-12}{F_4}^3 -
\frac{264}{5}\,(Da)^4a^{-5}F_4 +
\frac{192}{5}\,(Da)^3(DF_4)a^{-4} \\
&&+ 3\,(Da)^3\,a^{-4}F_3 + \frac{48}{5}\,(Da)^3\, a^{-9}{F_4}^2 +
\frac{9}{5}\,(Da)^2\,(DF_3) a^{-3} \\
&&- 12\,(Da)^2\,(D^2F_4)\,a^{-3} -
\frac{316}{25}\,(Da)^2\,(DF_4)\,a^{-8}F_4 +
\frac{24}{5}\,(Da)^2\,a^{-3}F_2 \\
&&- \frac{372}{25}\,(Da)^2\,a^{-8}F_3 F_4
+\frac{1056}{125}\,(Da)^2\,a^{-13}{F_4}^3  -
\frac{6}{5}\,(Da)\,(DF_2)\,a^{-2} \\
&&- \frac{6}{5}\,(Da)\,(D^2F_3)\,a^{-2} +
\frac{48}{25}\,(Da)\,(DF_3)\,a^{-7}F_4 +
2\,(Da)\,(D^3F_4)\,a^{-2}\\
&&+ \frac{64}{25}\,(Da)\,(D^2F_4)\,a^{-7} F_4
+\frac{44}{25}\,(Da)\,(DF_4)^2\,a^{-7} \\
&&+ \frac{68}{25}\,(Da)\,(DF_4)\,a^{-7}F_3 -
\frac{352}{125}\,(Da)\,(DF_4)\,a^{-12}{F_4}^2+ 2\,(Da)\,a^{-2}F_1
\\
&&- \frac{12}{5}\,(Da)\,a^{-7}F_2F_4 -
\frac{6}{5}\,(Da)\,a^{-7}{F_3}^2 +
\frac{66}{25}\,(Da)\,a^{-12}F_3{F_4}^2 \\
&&-\frac{88}{125}\,(Da)\,a^{-17}{F_4}^4 -
\frac{2}{5}\,(DF_1)\,a^{-1} + \frac{4}{25}\,(DF_2)\,a^{-6}F_4 \\
&&+ \frac{1}{5}\,(D^3F_3)\,a^{-1}
-\frac{6}{25}\,(DF_3)\,(DF_4)\,a^{-6} +
\frac{6}{25}\,(DF_3)\,a^{-6}F_3  \\
&&- \frac{18}{125}\,(DF_3)\,a^{-11}{F_4}^2 -
\frac{4}{25}\,(D^4F_4)\,a^{-1} - \frac{4}{25}\,(D^3F_4)\,a^{-6}F_4
\\
&&- \frac{8}{25}\,(D^2F_4)\,(DF_4)\,a^{-6} -
\frac{2}{25}\,(D^2F_4)\,a^{-6}F_3  +
\frac{4}{125}\,(D^2F_4)\,a^{-11}{F_4}^2 \\
&&+ \frac{24}{125}\,(DF_4)^2\,a^{-11}F_4 +
\frac{8}{25}(DF_4)a^{-6}F_2 - \frac{48}{125}\,(DF_4)a^{-11}F_3F_4  \\
 &&+ \frac{88}{625}(DF_4)a^{-16}{F_4}^3 + 2(D\sigma_{(-1)})
+ \frac{4}{5}\,a^{-1}F_0 + \frac{4}{25}\,a^{-1}\sigma_{(0)} \\
&&- \frac{4}{25}\,a^{-6} F_1F_4
 - \frac{4}{25}a^{-6} F_2F_3  + \frac{12}{125}\,a^{-11} F_2{F_4}^2 + \frac{12}{125}\,a^{-11}
{F_3}^2 F_4 \\
&&- \frac{44}{625}\,a^{-16} F_3{F_4}^3
+\frac{176}{15625}\,a^{-21}{F_4}^5,\qquad\qquad\qquad\qquad\qquad\quad\;\;\;(A.6)\\
\rho_{(5)}&=& 4\big(a^{-1}\sigma_{(1)} -
\rho_{(1)}\sigma_{(-1)}\big),\qquad\qquad\qquad\qquad\qquad\qquad\qquad\qquad\quad(A.7)
\end{eqnarray*}
where $F_{m} =\partial F/ \partial u_{m},\, m=1,\dots,5$ and
$\sigma_{(i)}, i=-1,0,1$ are differential polynomials such that
${(\rho_{(i)})}_{t}=D\sigma_{(i)}$. 

\end{document}